\documentclass[aps,prb,twocolumn,groupedaddress]{revtex4}

\usepackage{graphicx}% Include figure files
\usepackage{amsmath}
\usepackage{vmargin}

\setpapersize{USletter}
\setmarginsrb{1in}{1in}{1in}{1in}{0in}{0in}{0in}{0in}
\begin{document}

\title{First-principle studies of spin-electric coupling in a $\{Cu_3\}$ single molecular magnet}

\author{M. Fhokrul Islam}
\author{Javier F. Nossa}
\author{Carlo M. Canali}
\affiliation{School of Computer Science, Physics and Mathematics, Linnaeus University, Kalmar-Sweden}
\author{Mark Pederson}
\affiliation{Naval Research Laboratory, Washington, USA}
\date{\today}

\begin{abstract}
We report on a study of the electronic and magnetic properties of
the triangular antiferromagnetic $\{Cu_3\}$ single-molecule magnet,
based on spin density functional theory. Our calculations show
that the low-energy magnetic
properties are correctly described by an effective three-site spin $s=1/2$ Heisenberg model,
with an antiferromagnetic exchange coupling $J \approx 5$ meV. The ground state manifold of the
model is composed of two degenerate spin $S=1/2$ doublets of opposite chirality.
Due to lack of inversion symmetry in the molecule these two states are coupled
by an external electric field, even when spin-orbit interaction is absent.
The spin-electric coupling can be viewed as
originating from a modified exchange constant $\delta J$ induced by the electric field.
We find that the calculated transition rate between the chiral states yields
an effective electric dipole moment $d = 3.38\times 10^{-33} {\rm C\ m} \approx e 10^{-4}a$,
where $a$ is the Cu separation.
For external electric fields
${{\varepsilon}} \approx  10^8$ V/m this value corresponds to a Rabi time
$\tau \approx 1$ ns and
to a $\delta J$ of the order of a few $\mu$eV.
\end{abstract}


\maketitle

%%%%%%%%%%%%%%%%%%%%%%%%%%%%%%%%%%%%%%%%%%%%%%%%%%%%%%%%
%
%  INTRODUCTION
%
%%%%%%%%%%%%%%%%%%%%%%%%%%%%%%%%%%%%%%%%%%%%%%%%%%%%%%%%

\section{INTRODUCTION} \label{sec:INTRODUCTION}

Single-molecule magnets (SMMs) have been intensively studied in the
last two decades (for a review see Ref.~\onlinecite{Gatteschi2006}).
At low temperature these remarkable molecules behave in part like
bulk magnets thanks to their very long magnetization relaxation
time. At the same time SMMs are genuine quantum systems. They
display a variety of non-trivial quantum effects such as the quantum
tunneling of the magnetization,\cite{Friedman1996, Thomas1996} Berry
phase interference,\cite{Wernsdorfer1999} and quantum spin
coherence\cite{Ardavan2007}. Due to their double nature, SMMs are
ideal systems to investigate decoherence and the interplay between
classical and quantum behavior.\cite{Ardavan2007}

From the point of view of applications, interest in SMMs has been in
part spurred by the possibility that these structures could
represent the ultimate molecular-scale limit for magnetic units in
high-density magnetic storage materials. More recently SMMs have
been recognized as promising building blocks in molecular
spintronics, the emerging field combining spintronics and molecular
electronics.\cite{Rocha2005, Sanvito2006, Rocha2006, mhjo_nano06,
heersche_prl06, Bogani2008} In particular, thanks to their long spin
coherence time,\cite{Ardavan2007} SMMs are good candidates to
realize spintronic devices that  maintain, control and exploit
quantum coherence of individual spin states. These devices could
find important applications in the field of quantum information
processing.\cite{Leuenberger2001, Lehmann2007}

One key issue in using SMMs in molecular spintronics and quantum
information processing is the ability of switching efficiently
between their different magnetic states. The conventional way of
manipulating magnetic states is by applying an external magnetic
field. However, this approach has significant drawbacks when it
comes to controlling magnetic states at the molecular level. Quantum
manipulation of SMM requires application of an external field at a
very small spatial and temporal scale. It is, however, very
difficult to achieve such a small scale manipulation using standard
electron- spin control techniques such as electron spin resonance
(ESR) driven by ac magnetic field.\cite{Ardavan2007}

One promising alternative to achieve control of magnetic states at
the molecular level is to use an electric field instead. Typically,
by using STM tips for example, it is possible to apply strong
time-dependent electric fields in sub-nano regions, with time scales
of 1 ns.\cite{Hirjibehedin2006, Bleszynski-Jayich2008} Clearly,
since electric fields do not couple directly to spins, it is
essential to find efficient mechanisms for spin-electric coupling as
well as real SMMs where this mechanism can be at play. In principle
an electric field can interact with spins indirectly via the
spin-orbit interaction. However, since the strength of the coupling
scales like the volume of the system, this mechanism is not the
most efficient one for manipulating SMMs.

Recently, it has been proposed\cite{Trif2008} that in some molecular
antiferromagnets lacking inversion symmetry, such as the triangular
antiferromagnetic $\{Cu_3\}$ and other odd-spin rings, an electric
field can efficiently couple spin states through a combination of
exchange and chiralilty of the spin-manifold ground
state\cite{Trif2008}. The $\{Cu_3\}$ molecule, while
large\cite{Kortz2001, Choi2006}, reduces to a simple model composed
of three identical spin $s=1/2$ Cu cations coupled by an
antiferromagnetic (Heisenberg) exchange integration. Its ground
state consists of two total-spin $S=1/2$ doublets of opposite spin
chirality, degenerate in the absence of spin-orbit interaction.
According to an analysis based on group theory,\cite{Trif2008,
Trif2010} due to the lack of inversion symmetry, an electric field
can couple states of opposite chirality through the dipole operator,
even when spin-orbit interaction is absent. In the presence of an
additional small dc magnetic field that mixes the spin states, this
spin-electric coupling will then generate efficient electric
transitions from one spin state to another.

An intuitive picture of this coupling is the following. Since a spin
$S= 1/2$ triangular antiferromagnet is frustrated, there exist three
energetically degenerate antiferromagnetic spin configurations for
$S_z=1/2$  and three for $S_z=-1/2$. Both ground state chiral
eigenstates, with a given value of $S_z$, are appropriate, equally
weighted, linear combinations of these three frustrated spin
configurations. Each of these three configurations, if prepared,
would have a dipole moment with the same magnitude that points from
the antiparallel sites to the midpoint between the two parallel
sites. While the net dipole moment of the two chiral eigenstates is
zero, the dipole transition matrix element between them is not and
it is simply related to the magnitude of the permanent dipole moment
of the energetically degenerate frustrated configurations.

In practice the relevance of this spin-electric mechanism depends on
the coupling strength of the chiral states by the electric field,
{\it i.e.} on the value of the dipole moment of the frustrated spin
configurations. Theoretically this is an issue that only a
microscopic calculation for the specific molecule can address. The
main objective of this work is to calculate the strength of this
coupling for the $\{Cu_3\}$ molecule using {\it ab}-initio methods.
Our approach is based on Spin Density Functional Theory (SDFT),
implemented in the NRLMOL codes, which has been very successful in
describing the electronics and magnetic properties of
Mn$_{12}$-acetate and other SMMs.\cite{Pederson1999, Pederson2000,
Kortus2003, Postnikov2004} Recently SDFT implemented in NRLMOL has
been used in a first-principle study of quantum transport in a
Mn$_{12}$ single-electron transistor.\cite{michalak_prl_2010}

Our results show that indeed the crucial electric-dipole moment is
not negligible in $\{Cu_3\}$ and it would correspond to
characteristic Rabi times of 1 ns in the presence of typical
electric fields generated by STM tips. As originally suggested in
Ref.~\onlinecite{Trif2008}, the spin-electric coupling can be
interpreted as due to a modified exchange interaction brought about
by the electric field. Although here we only address the specific
case of $\{Cu_3\}$, our paper introduces a methodology that can be
followed in a systematic study of other SMMs without inversion
symmetry.

The paper is organized as follows. In section~\ref{cu3} we discuss
the electronic and magnetic  properties of triangular  $\{Cu_3\}$
molecule based on {\it ab}-initio calculations and show that the
low-energy quantum properties of the molecule can be described by an
effective three-spin $s= 1/2$ Heisenberg model with
antiferromagnetic coupling. In section~\ref{theory} we review the
underlying mechanism of spin-electric coupling in $\{Cu_3\}$
antiferromagnet, based on the effective spin Hamiltonian. The
first-principle computation of the spin-electric coupling  and
electric dipole moment of $\{Cu_3\}$ is presented in
section~\ref{d_E}. In section~\ref{J_E} we discuss the effect of the
electric field on the exchange coupling. Finally we present the
summary of our work in section~\ref{summary}.

%%%%%%%%%%%%%%%%%%%%%%%%%%%%%%%%%%%%%%%%%%%%%%%%%%%%%%%%
%
%   Electronic and magnetic properties of  $\{Cu_3\}$
%
%%%%%%%%%%%%%%%%%%%%%%%%%%%%%%%%%%%%%%%%%%%%%%%%%%%%%%%%

\section{Electronic and magnetic properties of  $\{Cu_3\}$}
\label{cu3}

%%%%%%%%%%%%%%%%%%%%%%%%%%%%%%%%%%%%%%%%%%%%%%%%%%%%%%%%
%
%  subsection   Microscopic description of the molecule
%
%%%%%%%%%%%%%%%%%%%%%%%%%%%%%%%%%%%%%%%%%%%%%%%%%%%%%%%%

\subsection{Microscopic description of the molecule}

The $\{Cu_3\}$ molecule that we are interested in has chemical
composition Na$_{12}$[Cu$_3$(AsW$_9$O$_{33}$)$_2\cdot$
3H$_2$O]$\cdot$32H$_2$O\cite{Kortz2001}. This molecule has been
studied experimentally by different groups.\cite{Kortz2001,
Choi2006} The three Cu$^{2+}$ cations form an equilateral triangle
and, as we show below, are the sites of three identical $s=1/2$
quantum spins. The frontier electrons on each of these sites have
primarily $d$ character. The bridging atoms consist of predominantly
paired electrons and are only polarized to the degree that the
same-spin states hybridize with the unpaired $d$-electrons on the Cu
sites. Due to the localized nature of transition-metal 3d states,
direct exchange stabilization due to parallel neighboring states is
expected to be exponentially small. Therefore, unless the frontier
$d$-electrons are spatially orthogonal by symmetry to the
$d$-electrons on other sites, {\it antiferromagnetic ordering}
between electrons on a pair of neighboring Cu atoms is energetically
preferred due to the increase in the system's kinetic energy,
induced by orthogonality constraints, when neighboring states are
parallel.

Although the spin model of three exchange-coupled spin $1/2$ is
quite useful to understand the magnetic properties of the $\{Cu_3\}$
SMM, all the other atoms in the molecule are essential for its
geometrical stability and for the resulting superexchange
interaction among the spins at the Cu sites. A proper {\it
ab}-initio description of the molecule must therefore include to a
certain extent all these atoms.

\begin{figure}[h]
\includegraphics[trim = 80mm 0mm 0mm 0mm, clip, scale=0.35]{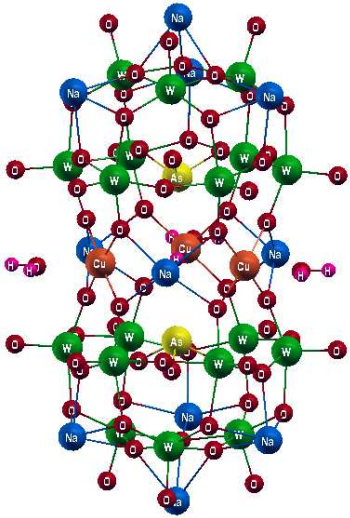}
\caption{Model of the $\{Cu_3\}$ molecule with chemical composition
Na$_{11}$H[Cu$_3$(AsW$_9$O$_{33}$)$_2\cdot$ 3H$_2$O] used in this
work. Xcrysden visualization tool\cite{Xcrysden} is used for this
figure.}
\label{mol}
\end{figure}

Building a suitable model of the molecule is a considerable
challenge since the model molecule should preserve the essential
physics. We have constructed the molecule by preserving the D$_{3h}$
symmetry of the polyanionic part of the molecule as observed in the
experiment.\cite{Kortz2001, Choi2006} Three of the twelve Na atoms
of the molecule are placed at the belt region of the molecule. These
three are the most important of all the Na atoms for the stability
of the belt region of the molecule. There is some uncertainty in the
position of the Na atoms but we have placed eight of the remaining
nine Na atoms in a way to preserve the D$_{3h}$ symmetry. The last
Na atom is replaced by a H atom and is placed at the center of the
molecule to maintain the charge neutrality of the valance electrons.
The model of the molecule used in this calculation is shown in
Fig.~\ref{mol}.

We have relaxed the geometry using the {\it ab}-initio package
NRLMOL\cite{Pederson1990, Jackson1990} that uses a Gaussian basis
set to solve the Kohn-Sham equations using PBE-GGA
approximation.\cite{Perdew} All-electron calculations are performed
for all elements of the molecule except for tungsten, for which we
have used pseudo potentials. The relaxation is first performed by
setting the net total spin of the molecule to $S= 3/2$ and then by
changing the net spin to $1/2$. Self-consistency is reached when the
total energy is converged to $10^{-6}$ Hartree or less.

\begin{figure}[h]
\includegraphics[trim = 12mm 0mm 0mm 0mm, clip, scale=0.26]{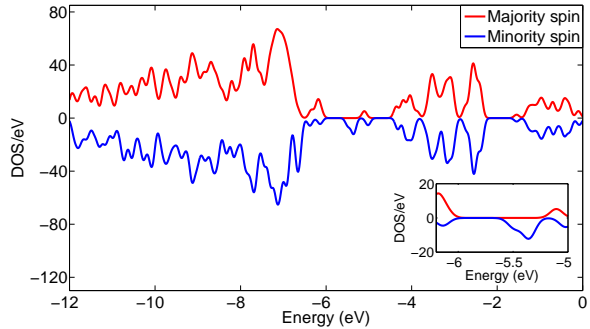}
\caption{Density of states of $\{Cu_3\}$ molecule. HOMO-LUMO gaps
for majority and minority spins are shown in the inset.}
\label{dos}
\end{figure}

The density of states of the molecule is shown in Fig.~\ref{dos}.
The HOMO-LUMO gap for the majority spin is calculated to be about
0.78 eV and that for minority spin is about 0.58 eV. Although in our
calculations we have used an equilateral arrangement of the three Cu
atoms, it is found experimentally that the $\{Cu_3\}$ molecule in
the ground state is slightly distorted into an isosceles
triangle.\cite{Kortz2001} Since the calculated HOMO-LUMO gap for the
equilateral configuration is relatively large, the distortion is
likely to be due to magnetic exchange rather than to the Jahn-Teller
effect.

One important result of our calculations, after the geometry
relaxation have been implemented, is that the ground state of the
system is anti-ferromagnetic,  with a net total spin of $1/2$ in
accordance with experiment.\cite{Stowe2004} The ground state energy
is lower by about 5.6 meV relative to spin $S= 3/2$ configuration.
This allows us to assign an exchange constant $J \approx $ 5 meV to
the three-site Heisenberg spin model mentioned above (see also next
Section).

The calculated magnetization density of the relaxed molecule shows
the presence of three electron-spin magnetic moments $\mu_i \approx
0.55 \mu_{\rm B},\  i = 1,2 ,3$, essentially localized at the three
Cu atom sites. Note that the orbital moments are quenched. These
results confirm that the low-energy properties of the $\{Cu_3\}$
molecule can be approximately described by an effective spin
Hamiltonian of three spins $s=1/2$ localized at the Cu sites.

\begin{figure}
\includegraphics[trim = 12mm 0mm 0mm 0mm, clip, scale=0.26]{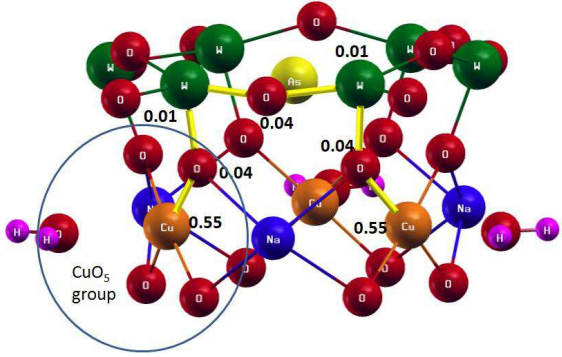}
\caption{(Color online) Superexchange coupling between two Cu atoms. The yellow
line connecting two Cu atoms through three O and two W atoms shows
the path along which spin coupling between Cu atoms is mediated. The
numbers near the atoms are the magnetic moment (in units of $\mu_B$)
of the atoms along the exchange path.}
\label{exchange}
\end{figure}

The exchange coupling between two Cu atoms is indirect and follows a
superexchange path\cite{Choi2006} along Cu-O-W-O-W-O-Cu as shown in
Fig.~\ref{exchange} - see the ye\-llow line connecting the atoms. To
understand this coupling mechanism we focus on one of the three
$CuO_5$ complexes of the molecule (shown inside the circle in
Fig.~\ref{exchange}). Because of the square-pyramidal $C_{4v}$
point-group symmetry of this complex, the $d_{xy}$, $d_{xz}$,
$d_{yz}$ states of Cu have lower energies compared to the
$d_{x^2-y^2}$ and $d_{z^2}$ states. Moreover, our calculation shows
that the axial Cu-O distance (2.35 A) in each unit is larger than
the four equatorial Cu-O distances (1.93 A). Thus the energy of
$d_{z^2}$ state is lower than $d_{x^2-y^2}$ state and the unpaired
{\it d} electron of the Cu$^{2+}$ ion resides in $d_{x^2-y^2}$ state
that is directed along the equatorial Cu-O vectors. Therefore, the
exchange coupling between two Cu atoms involves three O atoms and
two W atoms.

The magnetic moment calculations of the atoms of $\{Cu_3\}$ molecule
also support the superexchange path. The magnetic moments at the O
and W atoms on this path is much smaller than at the Cu sites, but
still 2 order of magnitude larger than at atoms not belonging to
this path.

%%%%%%%%%%%%%%%%%%%%%%%%%%%%%%%%%%%%%%%%%%%%%%%%%%%%%%%%
%
%  subsection   Effective spin Hamiltonian description
%
%%%%%%%%%%%%%%%%%%%%%%%%%%%%%%%%%%%%%%%%%%%%%%%%%%%%%%%%

\subsection{Effective spin Hamiltonian description}
\label{theory}

Based on the results of the {\it ab}-initio calculations, the low-energy
properties of the $\{Cu_3\}$ molecule can be described by the
following quantum spin Hamiltonian
\begin{equation}
H_0=\sum_{i=1}^3J_{i,i+1}{\bf s}_i\cdot{\bf
s}_{i+1}+\sum_{i=1}^3{\bf D}_{i,i+1}\cdot{\bf s}_i\times{\bf
s}_{i+1}
\label{ham}\,,
\end{equation}
where $J$ is the exchange parameter, ${\bf D}$ is the Dzyaloshinski
vector and ${\bf s}_i$ are three spins-1/2, located at the Cu sites.
The first term in the Hamiltonian is an isotropic Heisenberg model.
The geometry-relaxation and  electronic-structure calculations
showed that the Cu atoms form an equilateral triangle with a very
small intrinsic deformation. Since the atomic environment around
each of the the three Cu-Cu bonds is the same, we take the three
exchange constants $J_{i,i+1}$ to be the same value, $J$. On the
basis of the splitting between the ferromagnetic and
antiferromagnetic configurations discussed in the previous section,
$J$ is positive and $\simeq 5$ meV. The second term in Eq.
(\ref{ham}) is the anisotropic Dzyaloshinski-Moriya exchange
interaction origina\-ting from spin-orbit interaction. Its strength
$|{\bf D}_{i , i+1}|$ is at least one-order of magnitude smaller
than the isotropic exchange constant $J$, and we will disregard it
for the moment.

\begin{figure}
{\resizebox{3.2in}{2.4in}{\includegraphics{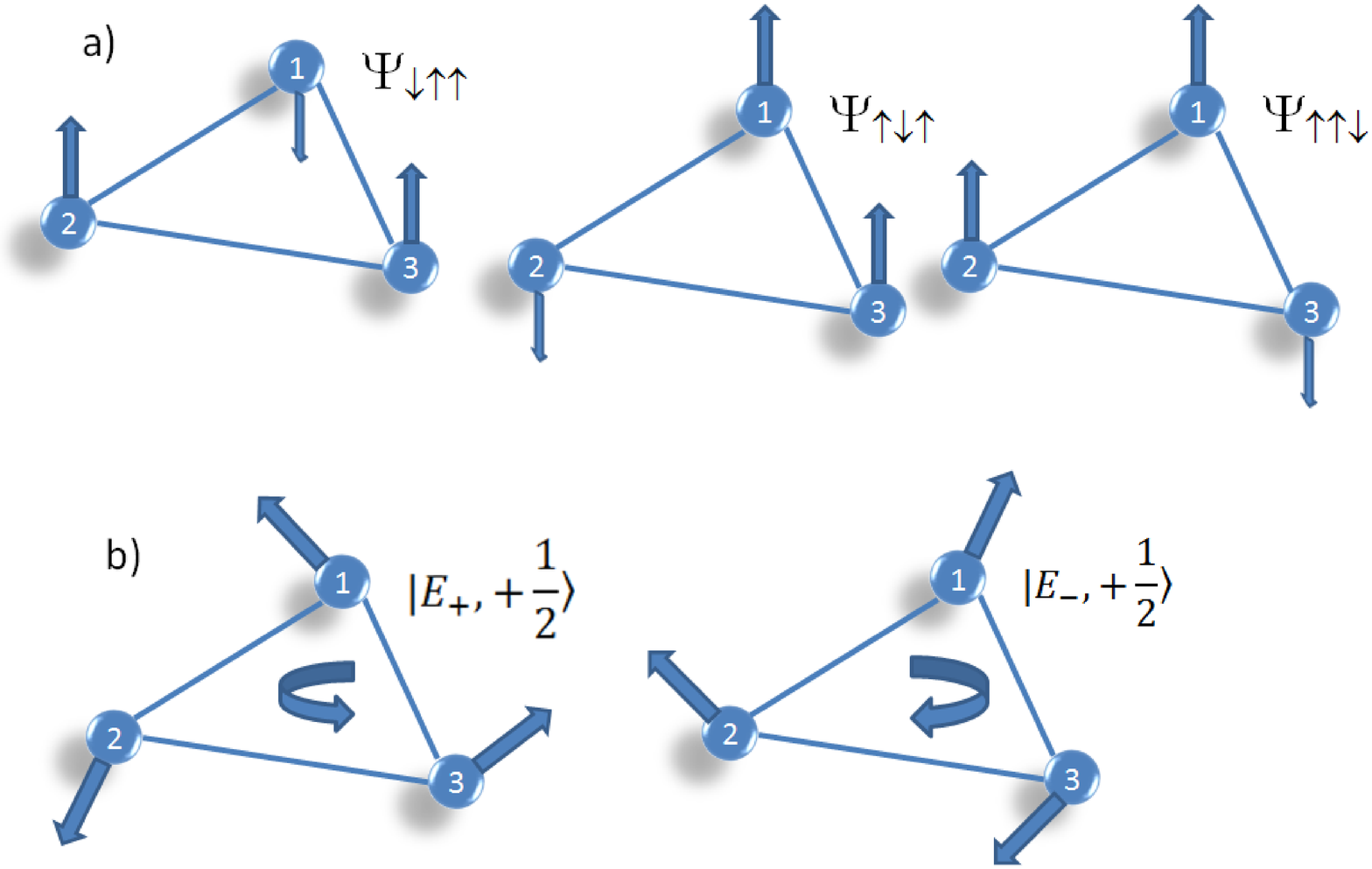}}}
\caption{a) The three spin configurations of the molecule associated
with total spin projection $S_z = +1/2$. b) The two chiral
states formed from a) with chirality +1 and -1, respectively.
\hfill\break } \label{state}
\end{figure}

The ground state of Eq.~(\ref{ham}) is  total spin $S=1/2$ manifold,
which can be constructed in terms of six degenerate spin
configurations, three associated with $S_z=+1/2$ and the other three
associated with $S_z=-1/2$. Fig.~\ref{state} shows the three
possible spin configurations associated with $S_z=+1/2$.

The total-spin $S=3/2$ four-dimensional subspace has an energy of
order $J$ above the ground-state manifold.

Within the $S=1/2$ ground-state manifold, we can construct two
degenerate, linearly independent doublets. Specifically the two
$S_z=+1/2$ states (shown in Fig.~\ref{state}(b)) are
%
%
%   chiral states
%
%
\begin{eqnarray}
|E_+,+\frac{1}{2}\rangle &=&
\frac{1}{\sqrt{3}}[\Psi_{\downarrow\uparrow\uparrow}+\omega\Psi_{\uparrow\downarrow\uparrow}+
\omega^2\Psi_{\uparrow\uparrow\downarrow}]\;, \nonumber\\
|E_-,+\frac{1}{2}\rangle &=&
\frac{1}{\sqrt{3}}[\Psi_{\downarrow\uparrow\uparrow}+\omega^2\Psi_{\uparrow\downarrow\uparrow}+
\omega\Psi_{\uparrow\uparrow\downarrow}]\;,
\label{chi1}
\end{eqnarray}
while the $S_z=-1/2$ states are
\begin{eqnarray}
|E_+,-\frac{1}{2}\rangle =
\frac{1}{\sqrt{3}}[\Psi_{\uparrow\downarrow\downarrow}+\omega\Psi_{\downarrow\uparrow\downarrow}+
\omega^2\Psi_{\downarrow\uparrow\uparrow}]\;, \nonumber \\
|E_-,-\frac{1}{2}\rangle =
\frac{1}{\sqrt{3}}[\Psi_{\uparrow\downarrow\downarrow}+\omega^2\Psi_{\downarrow\uparrow\downarrow}+
\omega\Psi_{\downarrow\downarrow\uparrow}]\;,
\label{chi2}
\end{eqnarray}
where $\omega = e^{\frac{i2\pi}{3}}$. The quantum numbers $E_+$ and $E_-$ specify the so called
handness or chirality of the states $|E_{\pm }, M\rangle$, which are eigenstates
of the chirality operator
\begin{equation}
C_z = \frac{4}{{\sqrt{3}}}{\bf s}_1\cdot {\bf s}_2\times {\bf s}_3\;,
\end{equation}
with eigenvalues $\pm 1$ respectively. It is useful to introduce
also the other two components of the chiral vector operator
\begin{equation}
C_x = -\frac{2}{3}({\bf s}_1\cdot {\bf s}_2 - 2 {\bf s}_2\cdot {\bf s}_3 + {\bf s}_3\cdot {\bf s}_1)\;,
\label{Cx}
\end{equation}
\begin{equation}
C_y = \frac{2}{{\sqrt{3}}} ({\bf s}_1\cdot {\bf s}_2 - {\bf s}_3\cdot {\bf s}_1)\;,
\label{Cy}
\end{equation}
and the ladder operators $C_{\pm} \equiv C_x \pm  i C_y$. Note that
$[C_l, C_m] = i 2 \epsilon_{lmn} C_n$ and $[C_l, S_m] = 0$. Here
$\epsilon_{lmn}$ is the Levi-Civita symbol. The ladder operators
reverse the chirality of the states: $C_{\pm} |E_{\mp}, M\rangle
=|E_{\pm}, M\rangle$. Thus $\bf C$ behaves exactly like the operator
$\bf S$ (for $S= 1/2$) in chiral space.

In the microscopic description of the molecule implemented
within density functional theory via the NRLMOL code,
the chiral states defined in Eqs.~(\ref{chi1}) and  (\ref{chi2})
have to be understood as being composed both of a spin and an orbital part.

We conclude this section with an observation of the DM interaction.
As shown in Ref.~\onlinecite{Trif2008}, the DM interaction within
the $S=1/2$ ground state manifold takes the simple form $H_{\rm DM}
=\Delta_{\rm SO} C_z S_z$, where $\Delta_{\rm SO}$ is the effective spin orbit coupling constant. Thus equal-spin states of opposite
chirality are split by $2\Delta_{\rm SO}$.

%%%%%%%%%%%%%%%%%%%%%%%%%%%%%%%%%%%%%%%%%%%%%%%%%%%%%%%%
%
%  section   Spin-electric effect in $\{Cu_3\}$
%
%%%%%%%%%%%%%%%%%%%%%%%%%%%%%%%%%%%%%%%%%%%%%%%%%%%%%%%%

\section{Spin-electric effect in $\{Cu_3\}$}
\label{spin_electric_effect}

%%%%%%%%%%%%%%%%%%%%%%%%%%%%%%%%%%%%%%%%%%%%%%%%%%%%%%%%
%
%  subsection   Absence of inversion symmetry and coupling of ground-state chiral states
%
%%%%%%%%%%%%%%%%%%%%%%%%%%%%%%%%%%%%%%%%%%%%%%%%%%%%%%%%

\subsection{Absence of inversion symmetry and coupling of ground-state chiral states}

The triangular spin-1/2 antiferromagnet $\{Cu_3\}$ belongs to the
class of antiferromagnetic rings with an odd number of half-integer
spins. In these systems, the lack of inversion symmetry of the
molecule as a whole implies that the ground-state is a
four-dimensional manifold, whose basis states $|E_{\pm}, S_z= \pm
1/2 \rangle$ are characterized by the spin projection $S_z = \pm
1/2$ and by the chirality $C_z = \pm 1$ (which we also label as
$E_{\pm}$ ). In contrast, antiferromagnetic rings with an even
number of spins have non-degenerate $S=0$ singlet ground state.
According to the original proposal in Ref.~\onlinecite{Trif2008,
Trif2010}, in odd-spin rings the two states of opposite chirality
$|E_{\pm}, S_z= M \rangle$ can be coupled linearly by an external
electric field, even in the absence of spin-orbit interaction. In
order for electric coupling to be non-zero, other criteria must be
satisfied.\cite{Trif2010} First of all, permanent electric dipoles
${\bf d}_{ij}$ must be present on the bridges that mediate the
coupling of spin ${\bf s}_i$ and ${\bf s}_j$. A necessary (although
not sufficient) condition for this is that the superexchange bridge
that magnetically couples ${\bf s}_i$ and ${\bf s}_j$ lacks a center
of inversion symmetry. Even when local dipole moments are present on
individual bridges, the resulting final spin-electric coupling
between chiral states depends in a nontrivial way on the overall
symmetry of the molecule. The best way to settle this issue is to
carry out a systematic symmetry analysis based on group theory. It
turns out that in triangular spin-1/2 antiferromagnets the coupling
is non-zero. On the other hand in pentagon spin 1/2
antiferromagnets, the coupling vanishes, unless spin-orbit
interaction is included.\cite{Trif2010}

We focus now on the spin-electric coupling of chiral states in
$\{Cu_3\}$. In the presence of an external electric field
$\pmb{\varepsilon}$, The Hamiltonian acquires the additional
electric-dipole term $H_{\varepsilon}=\sum_{i} e{\bf
r}_i\cdot{\pmb{\varepsilon}}=e{\bf R}\cdot\pmb{\varepsilon}$, where
$e$ is the electron charge and ${\bf r}_i$ is the coordinate of the
$i_{\rm th}$ electron.

In the subspace of spin projection $S_z = 1/2$ of the ground-state
manifold, which is invariant for  $H_{\varepsilon}$ the perturbed
Hamiltonian $H_0 + H_{\varepsilon}$ can be expressed in the basis of
the chiral states as
\begin{equation}
\begin{array}{l}
H=H_0+H_{\varepsilon} \\
=\left|
\begin{array}{cc}\langle E_+,+\frac{1}{2}|H_0|E_+,+ \frac{1}{2}\rangle &
\langle E_+, +\frac{1}{2}|H_{\varepsilon}|E_-, + \frac{1}{2}\rangle \\
   &  \\
\langle E_-, +\frac{1}{2}|H_{\varepsilon}|E_+, +\frac{1}{2}\rangle & \langle
E_-, +\frac{1}{2}|H_0|E_-, + \frac{1}{2}\rangle\end{array}\right|\;,
\end{array}
\label{perturb}
\end{equation}
A similar expression holds for the $S_z = -1/2$ subspace.
The eigenvalues of $H$ are
\begin{equation}
E_\frac{1}{2}^{\pm}(\pmb{\varepsilon})=E_\frac{1}{2}^{\pm
}(0)\pm|{\bf d\cdot\pmb{\varepsilon}}|\;,
\label{eigenenergy}
\end{equation}
with $E_\frac{1}{2}^{\pm }(0) = \langle
E_{\pm},+\frac{1}{2}|H_0|E_{\pm},+\frac{1}{2}\rangle$, and the
corresponding eigenstates

\begin{equation}
%|{\pm }, 1/2 \rangle (\pmb{\varepsilon})
\left | \chi_\frac{1}{2}^{\pm}(\pmb{\varepsilon}) \right \rangle
=\frac{1}{\sqrt{2}}\left(|E_+,+\frac{1}{2}\rangle\pm\frac{|{\bf
d\cdot\pmb{\varepsilon}}|} {\bf
d\cdot\pmb{\varepsilon}}|E_-,+\frac{1}{2}\rangle\right)\;.
\label{eigenstate}
\end{equation}
Here we have introduced the
electric dipole matrix element ${\bf d}$, which couples
states of opposite chirality (but with the same spin projection)
\begin{equation}
{\bf d}=\langle E_+,+\frac{1}{2}| e{\bf
R}|E_-,+\frac{1}{2}\rangle\;.
\label{dipole_me}
\end{equation}

For the specific example of $\{Cu_3\}$ molecule only the matrix
elements of $X$ and $Y$ components of $\bf R$ are nonzero and
\begin{equation}
 \langle E_+,+\frac{1}{2}| e{
X}|E_-,+\frac{1}{2}\rangle = i \langle E_+,+\frac{1}{2}| e{Y}|E_-,+\frac{1}{2}\rangle = {d\over \sqrt{2}}\;,
\label{d_comp}
\end{equation}
where $d\equiv |{\bf d}|$.

The matrix element in Eq.~(\ref{dipole_me}) is the key quantity in
the spin-electric coupling mechanism. Substituting the expressions
for the chiral states from Eqs.~(\ref{chi1}) and using the
orthogonality of spin states we obtain
\begin{eqnarray}
{\bf d}=\frac{1}{3}(\langle \Psi_{\downarrow\uparrow\uparrow}|e{\bf
R}|\Psi_{\downarrow\uparrow\uparrow}\rangle
+ \omega\langle \Psi_{\uparrow\downarrow\uparrow}|e{\bf R}|\Psi_{\uparrow\downarrow\uparrow}\rangle+ \nonumber \\
\omega^2\langle \Psi_{\uparrow\uparrow\downarrow}|e{\bf
R}|\Psi_{\uparrow\uparrow\downarrow}\rangle)\;.
\label{d}
\end{eqnarray}
Evaluating the dipole matrix element between two states of opposite
chirality is therefore equivalent to calculating the dipole moment
of each of the three spin configurations. This matrix element
determines the strength of spin-electric coupling and we are
primarily interested in calculating this quantity by {\it ab}-initio
methods.

Finally, note that all the matrix elements  of the electric dipole
operator $e{\bf R}$ are identically zero in the $S=3/2$ subspace.
This is obvious since $\langle \Psi_{\uparrow\uparrow\uparrow} | e
{\bf R} | \Psi_{\uparrow\uparrow\uparrow}\rangle$ and
$\frac{1}{3}(\langle \Psi_{\downarrow\uparrow\uparrow}|e{\bf
R}|\Psi_{\downarrow\uparrow\uparrow}\rangle + \langle
\Psi_{\uparrow\downarrow\uparrow}|e{\bf
R}|\Psi_{\uparrow\downarrow\uparrow}\rangle+ \langle
\Psi_{\uparrow\uparrow\downarrow}|e{\bf
R}|\Psi_{\uparrow\uparrow\downarrow}\rangle)$ are both zero by
symmetry. We will confirm this result by direct {\it ab}-initio
calculations.

%%%%%%%%%%%%%%%%%%%%%%%%%%%%%%%%%%%%%%%%%%%%%%%%%%%%%%%%
%
%  subsection   Effective spin Hamiltonian description
%
%%%%%%%%%%%%%%%%%%%%%%%%%%%%%%%%%%%%%%%%%%%%%%%%%%%%%%%%

\subsection{Effective spin Hamiltonian description}

The effect of the electric field on the the low-energy spectrum of
$\{Cu_3\}$ can be recast in the form of the effective spin model
introduced in Sec.~\ref{theory}. Since the electric dipole operator
has nonzero matrix elements only in the ground-state manifold, where
it couples states with equal spin components and opposite chirality,
we expect that the spin-electric Hamiltonian $H_{\varepsilon}$ can
be rewritten as a linear combination of the ladder operators
$C_{\pm}$. By comparing the matrix elements of $H_{\varepsilon}$
given in Eq.~(\ref{dipole_me}) and (\ref{d_comp}) with the action of
$C_{\pm}$ on the chiral states, one can show that\cite{Trif2010}
\begin{equation}
H^{\rm eff}_{\varepsilon} = {d\over \sqrt{2}} \pmb{\varepsilon}'\cdot {\bf C}_{\|}\;,
\end{equation}
where $\pmb{\varepsilon}' = R_z(\phi) (7 \pi/6 -2\theta )
\pmb{\varepsilon} $, with $R(\phi)$ being the matrix representing a
rotation by an angle $\phi$ around the $z$-axis, and $\theta$ being
the angle between the in-plane component $\pmb{\varepsilon}_{\|}$ of
the electric field and the bond ${\bf s}_1-{\bf s}_2$. By using
Eq.~(\ref{Cx}) and (\ref{Cy}) we can now rewrite ${\bf C}_{\|} =
(C_x, C_y)$ in term of spin-operators ${\bf s}_i$ and we
obtain\cite{Trif2010}
\begin{equation}
H^{\rm eff}_{\varepsilon} = \sum _i^3 \delta J_{ii+1}(\pmb{\varepsilon}) {\bf s}_i \cdot {\bf s}_j\;,
\label{h_e_eff}
\end{equation}
where the modified exchange parameters take the form \cite{Trif2010}
\begin{equation}
\delta J_{ii+1}(\pmb{\varepsilon})=\frac{4d}{3\sqrt{2}}|{\bf
\pmb{\varepsilon}}_{\|}|cos(\frac{2\pi}{3}i+\theta)\;.
\label{loss_J}
\end{equation}

This expression of the effective electric-dipole Hamiltonian
suggests a transparent physical interpretation of the spin-electric
couping mechanism.~\cite{Trif2008, Trif2010} An external electric
field changes the charge distribution of the $\{Cu_3\}$ molecule
which, in turn, changes the exchange interaction between neighboring
atoms. Since the modified exchange interaction does not commute with
$H_0$, it can cause transitions between chiral states within the
ground-state manifold.

In Eq. (\ref{loss_J}), $\pmb{\varepsilon}_{\|}$ is the projection of
electric field on the $Cu_3$ plane (in our case
$\pmb{\varepsilon}_{\|}=\pmb{\varepsilon}$), i=1 and $\theta=30^0$
is the angle between $\pmb{\varepsilon}$ and the line joining $Cu_1$
and $Cu_2$. Finally, note that Eqs.~(\ref{h_e_eff}) and
(\ref{loss_J}) provide an estimate of the dependence of the
ground-state energy as function of the electric field. Since in the
absence of spin-orbit coupling the electric-dipole Hamiltonian has
zero matrix elements in the $S=3/2$ subspace, Eq.~(\ref{loss_J})
gives us an estimate of the dependence of the exchange constant $J$
(proportional to the splitting between the $S=1/2$ ground state and
$S=3/2$ excited state) on $\pmb{\varepsilon}$.

%%%%%%%%%%%%%%%%%%%%%%%%%%%%%%%%%%%%%%%%%%%%%%%%%%%%%%%%
%
%  section   {\it ab}-initio evaluation of the spin-electric coupling
%
%%%%%%%%%%%%%%%%%%%%%%%%%%%%%%%%%%%%%%%%%%%%%%%%%%%%%%%%

\section{{\it ab}-initio evaluation of the spin-electric coupling}
\label{d_E}

%%%%%%%%%%%%%%%%%%%%%%%%%%%%%%%%%%%%%%%%%%%%%%%%%%%%%%%%
%
%  subsection   Calculation of the electric dipole moment
%
%%%%%%%%%%%%%%%%%%%%%%%%%%%%%%%%%%%%%%%%%%%%%%%%%%%%%%%%

\subsection{Calculation of the electric dipole moment}
To construct the chiral states of the full $\{Cu_3\}$ molecule, we
have calculated the ground state of the molecule for different spin
configurations, as shown in Fig.~\ref{state}. Although there are two
doublets of chiral states for the triangular arrangement of three
spin $1/2$ atoms, in this calculation we have used only one
doublet associated with the spin projection $+1/2$, since we are
interested in coupling between states of opposite chirality with
the same spin projection.

\begin{figure}[h]
\includegraphics[scale=0.4]{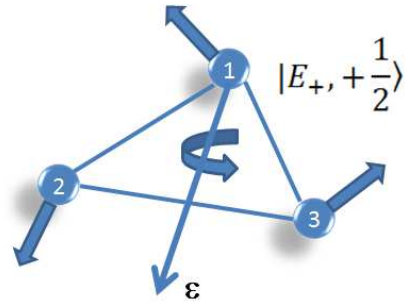}
\caption{The direction of the applied electric fields used in this
         calculation.}
\label{efield}
\end{figure}

To study the spin-electric effect we have applied an external field
along the perpendicular bisectors between positions 2 and 3 of the
$Cu_3$ triangle shown in Fig.~\ref{efield}, and have calculated the
corresponding ground state energy self-consistently for different
spin configurations. We have kept the direction of the field
relative to coordinate axes fixed, and have changed the orientation
of the spins at the Cu atoms to generate the three possible spin
configurations of the $\{Cu_3\}$ molecule.

\begin{figure}[h]
\includegraphics[scale=0.4]{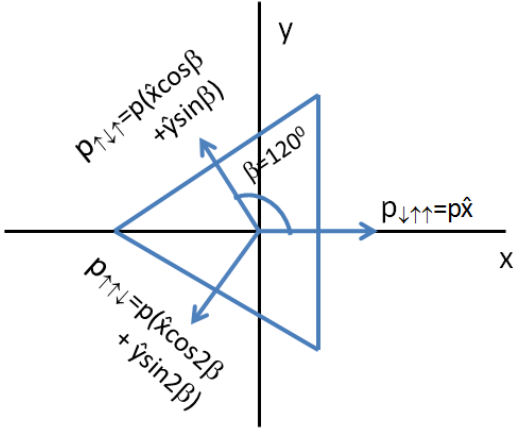}
\caption{Dipole moments of three spin configurations and their
relative angles. ${\bf
p}_{\downarrow\uparrow\uparrow}=\langle\Psi_{\downarrow\uparrow\uparrow}|e{\bf
R}|\Psi_{\downarrow\uparrow\uparrow}\rangle$, ${\bf
p}_{\uparrow\downarrow\uparrow}=\langle\Psi_{\uparrow\downarrow\uparrow}|e{\bf
R}|\Psi_{\uparrow\downarrow\uparrow}\rangle$ and ${\bf
p}_{\uparrow\uparrow\downarrow}=\langle\Psi_{\uparrow\uparrow\downarrow}|e{\bf
R}|\Psi_{\uparrow\uparrow\downarrow}\rangle$ are the moments
corresponding to the spin configurations of Fig.~\ref{state}a.}
\label{dipole}
\end{figure}

Our calculations show that $\{Cu_3\}$ molecule in the spin $S_z=3/2$
state does not have any permanent electric-dipole moment. On the
other hand each of the three frustrated spin $S_z=1/2$
configurations have a small permanent (i.e, zero-field) dipole
moment, as expected from the general discussion of Sec.~III. The
three moments have all the same magnitude but their directions are
along the perpendicular bisector of $Cu_3$ triangle and between two
Cu atoms with parallel spin alignments. The relative orientations of
these moments along with components are shown in Fig.~\ref{dipole}.
The fact that the $S_z=3/2$ state does not have permanent dipole
moment whereas $S_z=1/2$ states do, suggests that the dipole moments
are solely due to spin effects.

\begin{figure}[h]
\includegraphics[trim = 20mm 0mm 0mm 0mm, clip,scale=0.28]{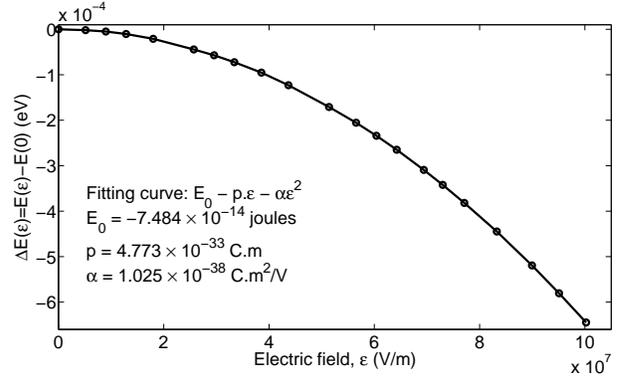}
\caption{Electric field dependence of the energy for one of the
three spin $S_z = 1/2$ spin configurations. The plot for the other
two configurations is very similar and the fitting yields
essentially the same values of $p$ and $\alpha$.}
\label{ene}
\end{figure}

In the presence of an electric field the energies of the $\{Cu_3\}$
molecule are slightly lower when field is between two Cu atoms with
parallel spins than for the other two spin configurations, where the
field is between two Cu atoms with anti-parallel spin alignments.
This difference in energy is due to the direction of permanent
moment relative to the induced moment. We have calculated the
permanent dipole-moment of the ground state spin configuration by
fitting the dependence of energy of one of the $S_z = 1/2$ spin
configurations with external field, as shown in Fig.~\ref{ene}. The
calculated values of the permanent dipole moment and polarizability
of $\{Cu_3\}$ molecule are $p=4.77 \times 10^{-33} C.m$ and
$\alpha=1.025 \times 10^{-38} C.m^2/V$, respectively. Although there
is no experimental value of polarizability available for $\{Cu_3\}$, 
polarizabilities within DFT calculations are generally accurate to
1-3 percent.

The value $p$ extracted from this fitting is consistent with the
direct calculation of the electric dipole moment of the three spin
configurations at zero field, implemented in the NRLMOL.

To calculate the matrix element ${\bf d}$ given in Eq.~(\ref{d}), we substitute the components of
the moments for the different spin configurations of Fig.~\ref{dipole},

\begin{eqnarray}
{\bf d}=\frac{1}{3}p[(1+\omega\cos\beta+\omega^2\cos2\beta){\bf\hat{x}} \nonumber \\
+(\omega\sin\beta+\omega^2\sin2\beta){\bf\hat{y}}] \nonumber \\
=\frac{p}{2}({\bf\hat{x}}+i{\bf\hat{y}})\;.
\label{df}
\end{eqnarray}

The magnitude of the dipole coupling in $\{Cu_3\}$ molecule is,
therefore
\begin{equation}
d=\frac{p}{\sqrt{2}}=3.38\times 10^{-33} C.m\;.
\label{d_cu3}
\end{equation}

The efficiency of the $\{Cu_3\}$ molecule as a switching device
depends on how fast an electric field can generate transitions from
one chiral state to the other. The characteristic (Rabi) time for
transitions between the two chiral states is given by
\begin{equation}
\tau = \frac{h}{|{\bf d}\cdot{\bf \pmb{\varepsilon}}|}\;.
\label{rabi}
\end{equation}

\begin{figure}[h]
\includegraphics[trim = 20mm 0mm 0mm 0mm, clip, scale=0.27]{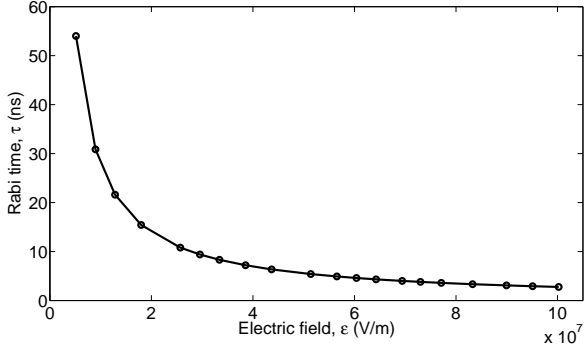}
\caption{Electric field dependence of the Rabi time for quantum transitions between the two (ground-state.}
\label{Rabi}
\end{figure}

Here, $h$ is Planck constant, ${\bf d}$ is the dipole matrix element between states of
different chirality given by Eq.~(\ref{d_cu3}), and
$\pmb{\varepsilon}$ is the external electric field. Fig.~\ref{Rabi}
shows the dependence of the Rabi time on external field, with the
maximum value of $\approx 50$ ns for a field $\varepsilon \simeq
5\times 10^6$  V/m. For larger fields of the order of $\simeq 10^8$
V/m, easily attainable in the vicinity of a STM tip, the Rabi time
is of the order of $1$ ns, which is considered to be a relatively
fast control-time in quantum information processing.

%%%%%%%%%%%%%%%%%%%%%%%%%%%%%%%%%%%%%%%%%%%%%%%%%%%%%%%%
%
%  subsection   Modification of exchange coupling in electric field
%
%%%%%%%%%%%%%%%%%%%%%%%%%%%%%%%%%%%%%%%%%%%%%%%%%%%%%%%%

\subsection{Modification of the exchange coupling in an electric field}
\label{J_E}

To calculate the dependence of the exchange coupling $J$ on the electric field,
we need to determine how the spin $S=1/2$ ground state and the
spin $S=3/2$  excited state depend on the field.
We define the exchange energy $J(\pmb{\varepsilon})$ as the difference
\begin{equation}
J(\pmb{\varepsilon})=E_{\frac{3}{2}}(\pmb{\varepsilon})-E_\frac{1}{2}^-(\pmb{\varepsilon})\;,
\label{ex_J}
\end{equation}
where $E_\frac{1}{2}(\pmb{\varepsilon})$ and $E_\frac{3}{2}(\pmb{\varepsilon})$ are
the energies of the $S=1/2$ ground-state and
of the spin $S=3/2$ excited state respectively in the presence of an electric field.

\begin{figure}[h]
\includegraphics[scale=0.28]{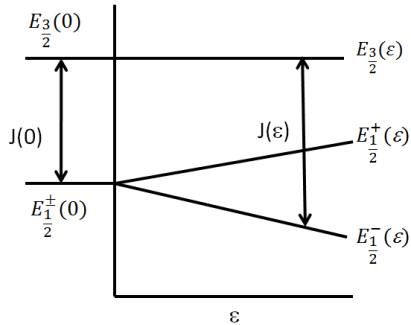}
\caption{Schematic electric-field dependence of the energies of the
$S=1/2$ chiral states and spin $S =3/2$ excited state, and the
exchange energy $J$ defined in Eq.~(\ref{ex_J}).} \label{chi_J}
\end{figure}

Based on on our discussion of
Sec.~\ref{spin_electric_effect}.A\  [see Eq.~(\ref{eigenenergy})], the energy of the
$S=1/2$ chiral ground-state manifold and the $S=3/2$ excited state vs $\varepsilon$
are shown schematically in Fig.~\ref{chi_J}, where we have
disregarded the quadratic dependence of both $E_\frac{3}{2}$ and
$E_\frac{1}{2}$ on the field due to the induced electric dipole moment.

The calculation of the electric-field-modified exchange parameter using first-principle methods is
not completely straightforward,
since the SDFT calculations done within NRLMOL allow us to calculate the energy of a
given spin configuration, whereas the (chiral) ground-state is a linear
combination of three possible spin configurations. However, we can
get an estimate of the dependence of $J$ on $\pmb{\varepsilon}$ by
approximating

\begin{eqnarray}
E_{\frac{1}{2}}^-(\varepsilon)&\approx&
\alpha^2_1\langle \Psi_{\downarrow\uparrow\uparrow}|H_{DFT}(\pmb{\varepsilon})|\Psi_{\downarrow\uparrow\uparrow}\rangle \nonumber \\
&&+\alpha^2_2\langle\Psi_{\uparrow\downarrow\uparrow}|H_{DFT}(\pmb{\varepsilon})|\Psi_{\uparrow\downarrow\uparrow}\rangle\nonumber \\
&&+\alpha^2_3\langle\Psi_{\uparrow\uparrow\downarrow}|H_{DFT}(\pmb{\varepsilon})|\Psi_{\uparrow\uparrow\downarrow}\rangle  \nonumber \\
&=&
 \alpha^2_1 E_{\downarrow\uparrow\uparrow}
+\alpha^2_2 E_{\uparrow\downarrow\uparrow}
+\alpha^2_3 E_{\uparrow\uparrow\downarrow}\;. \nonumber
\label{J_chi}
\end{eqnarray}
The coefficients $\alpha's$ can be obtained by expanding
$\left | \chi_\frac{1}{2}^{-}(\pmb{\varepsilon}) \right \rangle$ in
Eq.~(\ref{eigenstate}) in terms of the spin configurations, which leads
to
\begin{eqnarray}
\left | \chi_\frac{1}{2}^{-}(\pmb{\varepsilon}) \right \rangle&=&\frac{1}{\sqrt{6}}[(1-r)\Psi_{\downarrow\uparrow\uparrow}+(\omega-\omega^2r)\Psi_{\uparrow\downarrow\uparrow}  \nonumber \\
&& +(\omega^2-\omega r)\Psi_{\uparrow\uparrow\downarrow}] \nonumber\\
&=&\alpha_1\Psi_{\downarrow\uparrow\uparrow}+\alpha_2\Psi_{\uparrow\downarrow\uparrow}
         +\alpha_3\Psi_{\uparrow\uparrow\downarrow}\;,
\end{eqnarray}
where r=$\frac{|{\bf d}\cdot{\bf\pmb{\varepsilon}}|}{{\bf d}\cdot{\bf
\pmb{\varepsilon}}} =\frac{1}{\sqrt{2}}(1-i)$,
for the given choice of the electric field direction.

Therefore,
\begin{eqnarray}
E_{\frac{1}{2}}^-(\pmb{\varepsilon})&\approx&\frac{1}{6}\Bigg [\left(2-\sqrt{2}\right)E_{\downarrow\uparrow\uparrow}
+
\left(2 + {1 - \sqrt{3} \over \sqrt{2}}\right)E_{\uparrow\downarrow\uparrow}
\nonumber \\
&&+
%\left(2+\sqrt{\frac{3}{2}}+\sqrt{\frac{1}{2}}\right)E_{\uparrow\uparrow\downarrow}\Bigg] %\right ]
\left(2+ {1 + \sqrt{3} \over \sqrt{2}}\right)E_{\uparrow\uparrow\downarrow}\Bigg]\;. %\right ]
\label{J_chi2}
\end{eqnarray}

The energies $E_{\uparrow\downarrow\uparrow}$ and
$E_{\uparrow\uparrow\downarrow}$ are the same because of symmetry.
Since the difference between $E_{\downarrow\uparrow\uparrow}$ and
$E_{\uparrow\downarrow\uparrow}$ is very small and near the accuracy
limit of our calculations, we further approximate
$E_{\downarrow\uparrow\uparrow} \approx
E_{\uparrow\downarrow\uparrow}$.

The exchange parameter $J$ becomes
\begin{equation}
J(\pmb{\varepsilon})\approx
E_{\uparrow\uparrow\uparrow}(\pmb{\varepsilon})-E_{\downarrow\uparrow\uparrow}(\pmb{\varepsilon})\;,
\label{J_eq}
\end{equation}
with $E_{\uparrow\uparrow\uparrow}(\pmb{\varepsilon})\equiv E_{\frac{3}{2}}((\pmb{\varepsilon})$.

\begin{center}
\begin{figure}[h]
\includegraphics[trim = 20mm 0mm 0mm 10mm, clip, scale=.27]{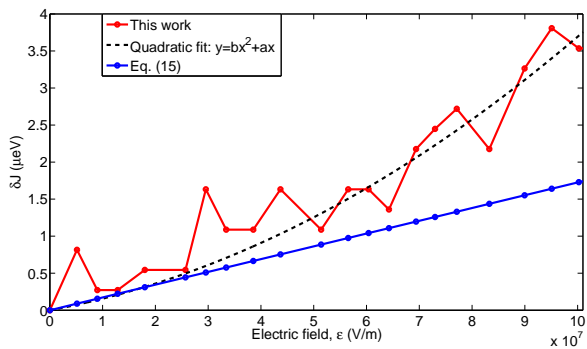}
\caption{Electric-field dependence of the variation of the exchange
energy $ \delta J(\varepsilon) \equiv J(\varepsilon) - J(0)$ induced
by the field. The red curve is the first-principle result obtained
by evaluating Eq.~(\ref{J_eq}) and the dashed black curve is the
quadratic fit of $\delta J(\varepsilon)$. The blue curve is a plot
of Eq.~(\ref{loss_J}) with the numerical value of $d$ extracted from
the first-principle calculations.} \label{J}
\end{figure}
\end{center}

In Fig.~\ref{J} we plot the electric-field-induced variation of the
exchange energy $\delta J(\varepsilon) \equiv J(\varepsilon) - J(0)
$ vs. $\varepsilon$. The result for $\delta J$ obtained by
evaluating Eq.~(\ref{J_eq}) with SDFT is shown by the red curve. For
this part of the calculations the convergence criterion has been
increased up to $10^{-8}$ Hartree. We can see that the dependence of
$J$ on electric field is quite small, and $\delta J$ is in the $\mu
e$V range for electric fields $\varepsilon = (1 - 10) \times 10^7$
V/m. These energies are not far form the accuracy limit of our
numerical calculations, which is the reason of the fluctuations seen
in the plot. Nevertheless the overall trend is an increase of
$\delta J(\varepsilon)$ with $\varepsilon$, which is approximately
linear at low fields. Note that the SDFT evaluations of
$E_{\uparrow\uparrow\uparrow}(\varepsilon)$ and
$E_{\downarrow\uparrow\uparrow}(\varepsilon)$ contain a quadratic
contribution in $\varepsilon$ but this nearly cancels at small
fields when computing $\delta J$, and it becomes appreciable only at
$\varepsilon \ge 5\times 10 ^7$ V/m.

The blue line in Fig.~\ref{J} shows the dependence of $\delta J$ on
$\pmb{\varepsilon}$ given by the prefactor of the cosine function in
Eq.~(\ref{loss_J}), which was derived within the spin Hamiltonian
formalism. When plotting Eq.~(\ref{loss_J}) we have used the value
of $d$ extracted from our first-principle calculations. Comparing
the two curves, we note that, apart form the fluctuations in the
numerical result mentioned above, the theoretical and numerical
values for $\delta J$ are consistent, and both procedures predict an
overall increase of $\delta J$ with electric field.

%%%%%%%%%%%%%%%%%%%%%%%%%%%%%%%%%%%%%%%%%%%%%%%%%%%%%%%%
%
%  section   Summary
%
%%%%%%%%%%%%%%%%%%%%%%%%%%%%%%%%%%%%%%%%%%%%%%%%%%%%%%%%

\section{Summary}
\label{summary}

In this paper we have carried out a first-principle study of the
spin-electric coupling in single-molecule magnets (SMMs) without
inversion symmetry. Specifically, we have analyzed the clear-cut
case of the $\{Cu_3\}$ triangular antiferromagnet where, because of
spin frustration, the ground-state consists of two generate spin
$1/2$ doublets of opposite chirality. Theory predicts\cite{Trif2008,
Trif2010} that an electric field can couple these states, even when
spin-orbit interaction is absent. The main goal of our work has been
to compute how strong this coupling is.

Our calculations of the electronic structure of the $\{Cu_3\}$
molecule show that the spin magnetic moments are localized at the
three Cu atom sites of the molecule. The magnetic properties of the
molecule are correctly described by a triangular spin $s=1/2$
Heisenberg antiferromagnet, with an exchange coupling $J$ of the
order of 5 meV that separates the energies of the spin-$S=1/2$
ground-state many-fold and the spin-$S=3/2$ excited states. In
agreement with theoretical predictions,\cite{Trif2008, Trif2010} we
find that an electric field couples the two ground-state doublets of
opposite chirality, even when spin-orbit interaction is absent. The
strength of the coupling is linear in the field and proportional to
the permanent electric dipole moment $d$ of the three frustrated
spin configurations. The calculations yield a value of $d \approx
4\times 10^{-33} {\rm C\ m} \approx e 10^{-4}a $ for $\{Cu_3\}$,
where $a$ is the Cu atom separation. Corresponding Rabi times for
electric-field-induced transitions between chiral states can be as
short as 1 ns, for electric fields of the order of $10^8$ V/m, which
are easily produced by a nearby STM tip. Thus this spin-electric
coupling mechanism is of potential interest for the use of
single-molecule magnets in quantum information processing as fast
switching devices.

Our calculations also indicate that the presence of an external
electric field modifies the exchange constant $J$. Typically the
electric field increases $J$, although the energy scale of this
change is in the $\mu e$V range for typical STM-generated electric
fields. Thus for this specific antiferromagnetic SMM, the electric
field cannot trigger directly a level crossing between magnetic
states with different total spin, as suggested recently for other
SMMs.\cite{sanvito_nat_mat_2009, van_der_zant_2010}

This work shows that a microscopic investigation of the
spin-electric coupling using the NRLMOL first-principle code is
feasible, and can systematically implemented for a large class of
SMMs which lack inversion symmetry. In this paper we have
disregarded the effect of spin-orbit interaction and external
magnetic field. The spin-orbit interaction strength is small
compared to the exchange coupling $J$. In the case of $\{Cu_3\}$ it
simply introduces a small splitting between the chiral states, but
is not expected to influence significantly the spin-electric
coupling. However in other antiferromagnetic rings with an odd
number of spins spin-orbit interaction is essential for the very
existence of the coupling mechanism.\cite{Trif2010}. Work to include
both spin-orbit interactions and an external magnetic field is in
progress. Together with the group-theory analysis presented in
Ref.~\onlinecite{Trif2010}, these studies will be a considerable
help in guiding future experiments and selecting the most promising
SMMs for applications in quantum information processing and
nanospintronics.

\section*{Acknowledgment}
We would like to thank Daniel Loss for introducing us to this
problem and for several useful discussions. This work was supported
by the Faculty of Natural Sciences at Linnaeus University, and the
Swedish Research Council under Grant No: 621-2007-5019.

\bibliography{Cu3biblio}

\end{document}